\documentclass[preprint,prd,noshowpacs]{revtex4}
\usepackage{amssymb}
\usepackage{amsmath}

\begin{document}

\title{Field localization and the Nambu--Jona-Lasinio mass generation mechanism in an 
alternative 5-dimensional brane model}

\author{Preston Jones}
\email{pjones@csufresno.edu}
\affiliation{Physics Department, California State University Fresno, 2345 East San Ramon Avenue M/S 37, Fresno, 
California 93740-8031}
\author{Gerardo Mu{\~n}oz}
\email{gerardom@csufresno.edu}
\affiliation{Physics Department, California State University Fresno, 2345 East San Ramon Avenue M/S 37, 
Fresno, California 93740-8031}
\author{Douglas Singleton}
\email{dougs@csufresno.edu}
\affiliation{Physics Department, California State University Fresno, 2345 East San Ramon Avenue M/S 37, 
Fresno, California 93740-8031 
and \\
Department of Physics, Faculty of Mathematics and Natural Sciences, 
Institut Teknologi Bandung, Jalan Ganesha 10 Bandung 40132, Indonesia}
\author{Triyanta}
\email{triyanta@fi.itb.ac.id}
\affiliation{Department of Physics, Faculty of Mathematics and Natural Sciences, 
Institut Teknologi Bandung, Jalan Ganesha 10 Bandung 40132, Indonesia}

\date{\today}

\begin{abstract}
We consider a 5-dimensional brane world model with a single brane which is distinct from 
the well known Randall-Sundrum model. We discuss the similarities and differences between our brane model and 
the Randall-Sundrum brane model. In particular we focus on the localization of 5D fields with different spins 
-- spin 0, spin 1/2, spin 1 -- to the brane, and a self-consistent mass generation mechanism. 
We find that the brane model studied here has different (and in some cases superior) localization properties for
fields/particles with different spins  to the brane, as compared to the original 5-dimensional 
brane models. In addition this alternative 5D brane model exhibits a self generation 
mechanism which recalls the self-consistent approach of Nambu and Jona-Lasinio.  
\end{abstract}


\maketitle

\section{Introduction}

One of the most active areas of recent research has been the work on large \cite{hamed1, hamed2, RaSu1} and infinite 
\cite{gogber1, gogber2, gogber3, gogber4, RaSu2} extra dimensions commonly referred to as ``brane" world models. 
A key motivation for these extra dimensional brane models is to address the problem of why the
energy scale of gravity (the Planck scale) should be 16 orders of magnitude larger than the 
energy scale of particle physics of the Standard Model (the TeV scale). A good feature of
these large and infinite extra dimensional theories is that they can be probed with
experiments that are in the reach of current technology. This is in distinction from traditional 
Kaluza-Klein theories or string theory where the direct, experimental
impact of the extra spatial dimensions is effectively non-existent at energy scales accessible with
current or reasonable extrapolations of near future technology. The possible experimental probes of these
large and infinite extra dimensional theories span the gamut from unique particle
accelerator signatures, such as the possible creation of mini-black holes at the LHC \cite{LHC-BH}, 
or deviations of Newton's inverse square law of gravity at micro-meter and smaller distances \cite{newton}.  

An important requirement of these theories with large and infinite extra dimensions is to explain why
at energy scales probed to the present the world appears to have only three spatial dimensions.
The answer given by these theories is that all matter particles/fields and all gauge particles/fields should
be bound to a 3+1 dimensional membrane (or ``brane" for short) in the higher dimensional 
space-time \cite{gogber3, gogber4}. In this way the world, as probed by matter particles of 
spin 0 or spin 1/2 or by force carrying gauge particles of spin 1, would still appear effectively 3+1 dimensional 
up to some energy scale. 

However, the original infinite extra dimensional brane metric of \cite{gogber1, gogber2, RaSu2} where not
able to localize all types of particles to the 3+1 brane in a simple manner. One could localize spin 0 fields 
on the brane but only at expense of not localizing spin 1/2 fields \cite{BaGa}. One could choose the 
parameters of the 5D metric \cite{gogber1, gogber2, RaSu2}, such as the sign of the of ``warp" factor, 
so as to localize the spin 1/2 fields but then 
spin 0 fields would not be localized \cite{BaGa}. And finally gauge bosons of spin 1 were not localized for
any choice of parameters of the 5D metric \cite{pomoral}. One could localize all the fields of various spin
in these 5D brane models by introducing additional non-gravitational interactions, but this spoiled the 
simplicity of the model.

In this paper we will investigate a simple variant of the 5D brane metrics of \cite{gogber1, gogber2, RaSu2} 
which appears similar to these original single-brane models. 
As far as we can find the fact that this alternative 5D warped metric is a different space-time has 
not been noted in the literature up to now nor have its properties been investigated. 
Additionally this alternative 5D metric has several key physical distinctions
with respect to the usual warped metric of \cite{gogber1, gogber2, RaSu2}: (i) It does
not require a fine-tuning between the brane tension and the bulk cosmological constant as is the
case with the usual 5D brane metric. (ii) It has different localization properties for
fields of different spins (i.e. spin 0, spin 1/2 and spin 1) to the brane. In particular we
find that for this alternative 5D brane metric
spin 1 fields can be localized with a decreasing warp factor. (iii) In the case of 
massive scalar fields localized to the brane one finds that the masses are generated by 
a self-consistent manner reminiscent of the mass generation mechanism of 
Nambu and Jona-Lasinio \cite{nambu}.

\section{Three 5D brane metrics for two 5D brane space-times}

Following \cite{gogber1, gogber2, RaSu1, RaSu2} we take the general gravitational action 
for the 5-dimensional brane world model, including a 5D cosmological constant, as
\begin{equation}
\label{grav-action}
S_g  =  - \frac{1}{{2 \kappa^2 }}\int {d^5 x {\sqrt g} R_5 }  + \frac{\lambda}{\kappa ^2} \int {d^5 x  {\sqrt g}  }  + S_{matter} ,
\end{equation}
where $R_5$ is the 5D Ricci scalar, $g$ is the determinant of the 5D metric, $\lambda$ is a 5D
cosmological constant and $S_{matter}$ is the action for any matter in the system. In the
Randall-Sundrum model \cite{gogber1, gogber2, RaSu1,RaSu2} $S_{matter}$ was a 4D
thin brane tension given by $\sigma \delta (z)$ i.e. a delta-function brane
tension $\sigma$. For the alternative 5D metrics which is the focus
of this paper we will find that $S_{matter}$ has, not only a delta-function
brane tension, but also a bulk energy-momentum.  (An excellent review of the 
standard brane models with the above type of action given in \eqref{grav-action}
can be found in \cite{Ru01}). To include some additional field, $\Phi$, one would expand the above 
action as $S=S_g + S_{\Phi}$ where $S_{\Phi}$ would be the action for the additional
scalar, spinor or gauge field. However, before studying the behavior of various matter and gauge fields in the 
5D space-time we will examine in some detail three 5D brane metrics. Two of the
metrics are just the Randall-Sundrum 1-brane metric. The third metric, which is the focus of 
this paper, at first appears to be some version of the Randall-Sundrum 1-brane metric will 
be shown to be different from the Randall-Sundrum 1-brane metric.      

We begin with the original form of the 5D brane metric \cite{gogber1, gogber2, RaSu1,RaSu2} which has the form
\begin{equation}
 \label{rsmetric}
  ds_{[y]} ^2 = e^{-2 k |y|}\eta_{\mu\nu}dx^\mu dx^\nu - dy^2 ,
\end{equation}
where $\eta_{\mu\nu} = diag (1, -1, -1, -1)$ is the flat 4D metric. 
The constant $k$ is fixed \cite{RaSu1,BaGa,Ru01} from the gravitational field equations with a fine tuning 
to the cosmological constant, $\lambda$. We will discuss this condition in the following section. 
An alternative version of the Randall-Sundrum model \cite{Muck} reverses 
the sign on the constant $k \to  - k$ and in the components of the stress-energy tensor.

A coordinate transformation for the extra dimension puts the metric in a more obviously conformally flat form. The coordinate
transformation to these conformally flat coordinates is $d y ^2  = e^{ - 2k\left|  y  \right|} dz^2$ \cite{Perez, Koyama,
Nima} and the metric and extrinsic curvature in the new system of coordinates is
\begin{equation}
\label{rsmetric1}
ds_{[z]} ^2  = \Omega ^2 (z) \eta _{AB } dx^A dx^B  = \frac{1}{{\left( {k|z| + 1} \right)^2 }}\eta _{AB } dx^A dx^B
\end{equation}
where the warp factor (i.e. $\Omega \left(z\right) = \frac{1} {{\left( {k|z| + 1} \right) }}$ no longer has 
an exponential form but is rather an inverse power of $|z|$. This $z$ coordinate system is a well known
alternative way to write the metric \eqref{rsmetric}. It is particularly useful in writing a Schr{\"o}dinger-like
equation for the gravitational perturbations of the 5D space-time. We will show in the following section 
that the two metrics \eqref{rsmetric} \eqref{rsmetric1} are alternative descriptions of the same space-time.

A third system of coordinates, which appears to be a hybrid of the $y$ system of \eqref{rsmetric}
and the $z$ coordinate system of \eqref{rsmetric1}, was suggested by various authors 
\cite{RaSu2,Karch,Davies07,Castro,Perez05,Perez,Csaki,Wolf}. This third system of coordinates, which we
unimaginatively call the $r$ coordinate system, is also
conformally flat but with an exponential warp factor rather than an inverse power of the coordinate
as in the case of the metric form \eqref{rsmetric1}. This metric has the form  
\begin{equation}
\label{rsmetric2}
ds _{[r]} ^2  = a ^2 (r) \eta _{AB } dX^A dX^B = e^{ -2 k |r|} \eta _{AB } dX^A dX^B 
\end{equation}
where the warp factor is again exponential $a \left( r \right) = e^{ -k |r|}$.  
There is a coordinate transformation which relates \eqref{rsmetric} with \eqref{rsmetric2} 
given by  $d y ^2  = e^{ - 2k\left| r \right|} dr^2  \to e^{ - k\left| r \right|}  = 1 - k\left| y  \right|$ 
and $e^{ - k|r|} dX^\mu   = e^{ - k| y |} dx^\mu   \to dX^\mu   = \frac{{e^{ - k\left| y  \right|} }}
{{1 - k\left| y  \right|}}dx^\mu$. Since there is a transformation relating the metrics in 
\eqref{rsmetric} and \eqref{rsmetric2}  one might be tempted to conclude that (as is the case for
metrics \eqref{rsmetric} and \eqref{rsmetric1}) the two metrics are the same. However, one hint that 
\eqref{rsmetric} and \eqref{rsmetric2} are different comes from the fact that the transformation 
which relates the two metrics is singular at $|y| = 1/k$. Further for
the transformation relating the $y$ coordinate system with the $r$ coordinate to be global 
one would need to require that the differential be exact,  $dX^\mu   = A(x_\mu,y) dx_\mu   + B(x_\mu,y) dy$ 
and that the derivative obey $\partial_y A = \partial_{x_\mu} B = 0$. This condition is not satisfied in general 
since $\partial_y A \ne 0$.  The conclusion is that the transformation 
between $y$ and $r$ coordinates is local but not global. 
However, on any bulk coordinate foliation with $y=y_c$ a constant $\partial_y A = 0 $ and the coordinate 
transformation is global. On the other hand for the coordinate transformation between
the $y$ and $z$ coordinates is global since one can obtain an exact differential. 
In the following section we will find other ways in which the 5D brane 
space-time given by \eqref{rsmetric2} is different from the original 5D brane metrics of
\eqref{rsmetric} and \eqref{rsmetric1}. We will also high light some important physical distinctions between
the different 5D brane world metrics. In particular the localization properties of 
the space-time given by the $r$ coordinate is different than the space-time given by
the $y$ and $z$ coordinate metrics. In addition the $r$-metric has a self consistent mass generation
mechanism given in \cite{nambu}. 

\section{Einstein equations and the stress energy tensors}

We now study the energy-momentum tensor of the metrics \eqref{rsmetric}, \eqref{rsmetric1}, and \eqref{rsmetric2}
by feeding them through the 5D Einstein equations. To facilitate the study of the three metrics we will write the
metrics in a generic form as
\begin{equation}
\label{gen-metric}
ds^2  = a^2  (| x^5 |)\eta _{\mu \nu } dx^\mu  dx^\nu   - b^2 ( |x^5 |)dx^5 dx^5 
\end{equation}
where $x_5 = r, y, z$ depending on whether one is dealing with metric \eqref{rsmetric2}, \eqref{rsmetric}
or \eqref{rsmetric1}. The ansatz functions $a, b$ are functions of the absolute value of the
fifth coordinate. The ansatz functions $a, b$ can be changed depending on which of the three metrics one is
dealing with (e.g. for the metric \eqref{rsmetric2} $a = b = e^{-k |r |}$).
 
We first calculate the Ricci scalar for the three metrics. In this way we can see that \eqref{rsmetric}
and \eqref{rsmetric1} are the same space-time, while \eqref{rsmetric2} is a different space-time. For 
the general ansatz \eqref{gen-metric} the Ricci scalar is
\begin{equation}
\label{Ricci}
R = 4\frac{{2aa'b' - 3a'^2 b - 4\delta \left( | x^5 | \right)aa'b - 2aa''b}}{{a^2 b^3 }}~,
\end{equation}
where the prime means differentiation with respect to $x_5$. 
By applying \eqref{Ricci} to each of the three metrics above we find
\begin{subequations}
\begin{equation}
\label{ricciscalary}
   {R _{\left[ y \right]}   =  - 20k^2  + 16k\delta \left( y  \right),}
\end{equation}
\begin{equation}
\label{ricciscalarz}
   {R _{\left[ z \right]}  =  - 20k^2  + 16k\delta \left( z \right) + 16k^2 \delta \left( z \right)z,}
\end{equation}
\begin{equation}
\label{ricciscalarx}
   {R _{\left[ r \right]}  =  - 12k^2 e^{2kr}  + 16k\delta \left( r \right)e^{2kr} }.
\end{equation}
\end{subequations}
The three different metrics/coordinate systems are indicated via the bracketed subscripts. Since 
$\delta \left( z \right)z=0$ one can see that the Ricci scalars for the $y$ and $z$ coordinates
are the same, indicating that the $y$ and $z$ metrics represent the same space-time. However 
$R _{\left[ r \right]}$ is different than $R _{\left[ y \right]}$ and $R _{\left[ z \right]}$ 
indicating that the metric \eqref{rsmetric2} is a different space-time than the space-time
given by the metrics \eqref{rsmetric} or \eqref{rsmetric1}.  

Next, we obtain the energy-momentum tensor connected with the metrics \eqref{rsmetric},
\eqref{rsmetric1}, and \eqref{rsmetric2} by feeding them through the 5D Einstein equations
\begin{equation}
\label{5d-einstein}
G_{AB}  + g_{AB} \lambda  = \kappa^2 T_{AB}~.
\end{equation}
Since all three metrics are symmetric about the location of the brane at $x_5 =0$ we will only focus 
on one side of the brane, namely $x_5 >0$ (this avoids the unnecessary complication of writing
the down step functions whenever there is a first derivative of $a (| x_5 |)$ and/or
$b (| x_5 |)$. The second derivatives of $a (| x_5 |)$ and/or $b (| x_5 |)$ will give
rise to $\delta (| x_5 |)$ i.e. the brane energy density. The components of the
Einstein tensor for the brane coordinates ($\mu , \nu = 0, 1, 2, 3$)
for the general metric \eqref{gen-metric} are 
\begin{equation}
G_{\mu \nu } = \frac{3}{{b^3 }}\eta _{\mu \nu } \left( {a'^2 b + 2aa'b\delta \left( x^5 \right) + aa''b - aa'b'} \right) ~,
\end{equation}
For the bulk coordinates one finds
\begin{equation}
G_{55}  =- 6\left( {\frac{{a'}}{a}} \right)^2 ~.
\end{equation}
In terms of the three coordinate systems -- $y$, $z$, and $r$ -- the Einstein equations on the brane become
\begin{subequations}
\begin{equation}
\label{uvy}
   {\eta _{\mu \nu } e^{ - 2k\left| y \right|} \left( {6k^2  + \lambda _{ [ y ]} } \right) - 6k\eta _{\mu \nu } 
\delta \left( y \right) = \kappa^2 T_{\mu \nu } ,}
\end{equation}
\begin{equation}
\label{uvz}
   {\eta _{\mu \nu } \frac{1}{{\left( {k\left| z \right| + 1} \right)^2 }}\left( {6k^2  + \lambda _{[ z ]} } 
\right) - 6k\eta _{\mu \nu } \delta \left( z \right) = \kappa^2 T_{\mu \nu } ,}
\end{equation}
\begin{equation}
\label{uvx}
   {\eta _{\mu \nu } \left( {3k^2  + e^{ - 2k\left| r \right|} \lambda _{[ r ]} } \right) - 6k\eta _{\mu \nu } 
\delta \left( r \right) = \kappa^2 T_{\mu \nu } ,}  \\
\end{equation}
\end{subequations}
where $g_{\mu \nu } \delta \left( y  \right)=\eta_{\mu \nu } \delta \left( y  \right)$ and $g_{\mu \nu } \delta 
\left( z \right)=\eta_{\mu \nu } \delta \left( z \right)$. These delta functions in the 4D
energy-momentum tensor indicate that the matter sources are thin branes. Such thin branes can be
obtained as the limit of ``thick" brane solutions \cite{bronn03} \cite{bronn07} \cite{dzh08}
\cite{dzh10}. The bulk Einstein equations for the three systems are,
\begin{subequations}
\begin{equation}
\label{55y}
   { - \left( {6k^2  + \lambda _{[ y ]} } \right) = \kappa^2 T_{55} ,}
\end{equation}
\begin{equation}
\label{55z}
   -{\frac{1}{{\left( {k\left| z \right| + 1} \right)^2 }}\left( {6k^2  + \lambda _{[ z ]} } \right) = \kappa^2 
T_{55} ,}
\end{equation}
\begin{equation}
\label{55x}
   { - 6k^2  -  e^{ - 2k\left| r \right|} \lambda _{[ r ]}  = \kappa^2 T_{55} .}
\end{equation}
\end{subequations}
From \eqref{uvy}, \eqref{uvz}, \eqref{55y}, and \eqref{55z} one notices that it is possible to reduce the
energy-momentum tensor to almost vacuum, with the exception of the non-zero brane tension
$- 6k\eta _{\mu \nu } \delta \left( y \right) / \kappa^2$ or
$- 6k\eta _{\mu \nu } \delta \left( z \right) / \kappa^2$, by fine tuning the constant $k$ to the 5D
cosmological constant via the condition $\lambda _{[ y ]}  = \lambda _{[z ]}   =  - 6k^2$.  Looking at 
the energy-momentum components for the $r$ coordinate system -- \eqref{uvx} and \eqref{55x} -- one can
see that a similar fine-tuning is not possible in this case. The simplest choice for this metric is to 
take the 5D cosmological constant as vanishing i.e.  $\lambda _{[ r ]} =  0$. With this choice the
energy momentum tensor for the $r$ coordinate system becomes
\begin{subequations}
\begin{equation}
\label{tuv-r}
  3k \eta _{\mu \nu } \left( k  - 2 \delta ( | r | ) \right)  = \kappa^2 T_{\mu \nu } ,
\end{equation}
\begin{equation}
\label{t55-r}
   { - 6k^2  = \kappa^2 T_{55} .}
\end{equation}
\end{subequations}
Thus the fifth component of the energy-momentum is a bulk constant $T_{55} = -6 k^2 / \kappa^2$.
The other components of the energy-momentum tensor , $T_{\mu \nu}$, are composed of a bulk 
constant term, $3k^2$, a constant term, $-  6 k \delta ( | r |)$ which, by the delta function,
is confined to the brane. For observers confined to the brane the effective
4D energy-momentum tensor will appear as an effective 4D 
cosmological constant term (i.e. $ T ^{4D} _{\mu \nu} = \Lambda_{4D} \eta _{\mu \nu}$
with $\Lambda _{4D} = 3k^2 - 6 k$). The sign of this effective 4D 
cosmological constant, $\Lambda _{4D}$ can be negative (for $0 < k < 2$) or positive 
(for $k<0$ or $k>2$). For $k=2$ effective 4D cosmological constant vanishes, but one
still has a constant bulk term. The difference in the energy-momentum tensors for the 
three metrics again indicates that the $y$ and $z$ metrics represent the same space-time, while
the $r$ metric is a related, but different space-time. 

The energy-momentum tensor in \eqref{tuv-r} \eqref{t55-r} can be split into a constant term on the
4D brane plus a constant part in the 5D bulk as follows
\begin{eqnarray}
\label{TAB}
T_{AB} &=& T_{AB} ^{brane} + T_{AB} ^{bulk} \nonumber \\
&=& \frac{-6 k \delta (r)}{\kappa^2} diag[1, -1, -1, -1, 0]
+ \frac{3k^2}{\kappa^2} diag[1, -1, -1, -1, -2] ~,
\end{eqnarray}
where $diag$ indicates a $5 \times 5$ diagonal matrix. The first term, proportional to $\delta (r)$, represents
a constant 4D energy-momentum tensor which is confined to the brane. The second term, proportional to 
$\frac{3k^2}{\kappa^2}$, represents a constant term in the 5D bulk, but with the complication that the
``pressure" term in the $rr$ direction is twice that of the other three spatial directions. This 
difference between the pressures in the three spatial coordinates of the brane and the bulk
spatial dimension could have been anticipated since there should be some difference between
the nature of energy densities and pressures of the matter sources on the brane and in the bulk.
If the matter sources had the same energy densities and pressures on the brane as in the bulk
there would be no difference between the brane and bulk, and one would not have a warped brane geometry.

We now briefly discuss what kind of field sources could 
give rise to the split, constant energy-momentum tensor as in \eqref{TAB}.
The most obvious choice is that the $T_{AB}$ could be generated by the condensate of some field(s), which for
simplicity we will take to be scalar fields. In regard to the first term in \eqref{TAB}, 
$T_{AB} ^{brane} = \frac{-6 k \delta (r)}{\kappa^2} diag[1, -1, -1, -1, 0]$, one can obtain such
an energy momentum tensor from a scalar field condensate like the Standard Model Higgs, which is confined
to the brane. In a realistic model this scalar field and its condensate would be confined to some
finite thickness region near $r=0$ rather than an infinitesimal thin region implied by $\delta (r)$.  
An energy-momentum tensor having the general form $T_{AB} ^{brane} = F(x) \delta (r)diag[1, -1, -1, -1, 0]$ 
(where $F(x)$ is some function of the 4D coordinates of the brane) can be obtained via any field and field 
condensate which is localized to the brane \cite{dvali}. In our case above $F(x) = const.$ 
Obtaining a field theory source which gives $T_{AB} ^{bulk}$ from \eqref{TAB} requires a bit
more thought due to the fact that the $rr$ component is twice that of the other spatial components.
An energy-momentum tensor of the form $T_{AB} ^{bulk}$ can be obtained from a ghost, scalar field, 
$\phi (x, r)$, having some self-interaction $V (\phi)$. The Lagrangian is of the form
\begin{equation}
\label{5D-Higgs}
{\cal L} _{5D} = - \frac{1}{2}(\partial _A \phi ) (\partial ^A \phi ) - V (\phi) ~.
\end{equation}
The unusual negative sign in front of the kinetic term indicates that this is a ghost field. While such 
fields are problematic, it has been argued \cite{koley, koley1, koley2} that this can be handled as long 
as the ghost fields are bulk fields (as is the case here). Also, effective ghost fields can arise in a natural 
way in the context of Weyl gravity  \cite{gog-sing-10}. Here our aim is simply to
find some field theory source which can give the energy-momentum tensor associated with the brane metric
of \eqref{rsmetric2}. The energy-momentum tensor associated with \eqref{5D-Higgs} is
\begin{equation}
\label{5D-T-P}
T_{AB} ^{\phi} = \frac{\partial {\cal L} _{5D}}{\partial \phi ^{,A}} \phi _{,B} - g_{AB} {\cal L} _{5D} ~,
\end{equation}
where as usual $\phi _{,A} = \partial _A \phi$. Using ${\cal L} _{5D}$ from \eqref{5D-Higgs} as 
well as the 5D metric \eqref{rsmetric2} in \eqref{5D-T-P} gives
\begin{eqnarray}
\label{5D-T-P-2}
T_{\mu \nu} &=& \eta _{\mu \nu} \left( - \frac{1}{2} (\partial _r \phi )^2  +  e ^{-2 k |r|} V (\phi) \right) \\
T_{rr} &=& \left( - \frac{1}{2} (\partial _r \phi )^2  - e ^{-2 k |r|} V (\phi) \right) ~.
\end{eqnarray}
The above forms for $T_{\mu \nu}$ and $T_{rr}$ take into account the fact that equations \eqref{uvx} 
\eqref{55x} and \eqref{5D-T-P} allow at most an $r$-dependent field. 
In order to obtain some energy-momentum tensor of the form $T_{AB} ^{bulk}$ from \eqref{TAB} we require
that $(\partial _r \phi )^2  = \frac{3k^2}{\kappa^2}$ and $e ^{-2 k |r|} V (\phi) 
= \frac{9 k^2}{2 \kappa^2}$ as one moves into the bulk i.e. away from $r=0$. The condition
$(\partial _r \phi )^2 = \frac{3k^2}{\kappa^2}$ 
can be met by $\phi (r) = \frac{\sqrt{3}k}{\kappa} r$. Note that this form of the ghost field
$\phi$ implies that it vanishes on the brane $r=0$. The condition 
$e ^{-2 k |r|} V (\phi) =  \frac{9 k^2}{2 \kappa^2}$ can be met by 
taking the potential to be of the appropriate form. Taking into account the behavior
of $\phi (r)$ off the brane, i.e. $\phi (r) = \frac{\sqrt{3}k}{\kappa}r$, the
potential has the form $V (\phi) =  \frac{9 k^2}{2 \kappa^2} \exp [2 \kappa \phi / \sqrt{3}]$.
Such exponential potentials are called Liouville potentials and arise in string theory 
as well as in some quintessence models \cite{quintessence}
\cite{quintessence1} \cite{quintessence2}. It is also possible to show that the above scalar field
solution and potential solve the field equation for the scalar field namely
$$
- \frac{1}{\sqrt{g}} \partial _A \left( \sqrt{g} g^{AB} \partial _B \phi \right) = -\frac{\partial V}{\partial \phi} ~, 
$$
where $g = e^{-10 k r}$ is the determinant of the metric. The linear character of the 
scalar field solution, i.e. $\phi \propto r$, is reminiscent of the linear potential which
is postulated to lead to the confinement of quarks. In a similar way one might think to use this
scalar field solution to localize other fields to the brane by coupling them to
$\phi$. However coupling ordinary matter/fields to a ghost field can lead to problems. Thus
we avoid coupling this ghost field directly to any ordinary fields. 

Thus it is possible to construct a field theory source (albeit with a ghost scalar field) 
that gives the energy-momentum tensor \eqref{TAB} associated with the brane metric
\eqref{rsmetric2}. Before leaving this topic of possible
field theory sources that might give an energy-momentum tensor of the form in \eqref{TAB} we recall
that in reference \cite{visser} a brane model was given with the ``less" warped metric of
\begin{equation}
\label{visser}
ds^2 = e^{2 f(r)} dt^2 - dx_i dx^i - dr^2 ~.
\end{equation}
``Less" warped means that the warp factor $e^{2 f(r)}$ sits only in front of the time component of the
metric; in the Randall-Sundrum metric \eqref{rsmetric} the warp factor sits in front of all the coordinates
except the extra spatial dimension; in the alternative brane metric \eqref{rsmetric2} the warp factor
sits in front of all the coordinates including the extra spatial dimension. For the metric in \eqref{visser}
it is possible to give an exact field theory sources in terms of a 5D $U(1)$ gauge boson with a
vector potential of the form $A^B (x ) = (a(r), 0, 0, 0, 0)$. This form of the 5D vector
potential in combination with the metric \eqref{visser} was shown to solve the 5D ``Maxwell"
equations and yield an energy-momentum tensor of the form $T_{AB} = K [1, -1, -1, -1, 1]$ where
$K$ is some constant. While not precisely of the form required to support the alternative 5D metric
\eqref{rsmetric2}, this energy-momentum tensor does have the feature in common with the energy-momentum
tensor of \eqref{TAB} that the pressure in the extra spatial direction, $r$, is different from that in the
three spatial dimensions on the brane. Thus in addition to the example
of the ghost scalar field theory source in \eqref{5D-Higgs} that would lead to the energy-momentum
tensor in \eqref{TAB} it might also be possible to expand on the simple $U(1)$ Abelian gauge source 
of \cite{visser} to find some non-ghost source which would yield \eqref{TAB}. For example one
might try some combination of 5D $U(1)$ gauge field coupled to regular scalar field, or one could
try a 5D non-Abelian gauge field or a 5D Born-Infeld $U(1)$ field.

Before moving on to the localization of fields onto the brane, located at $r=0$ for the 
$r$-metric, we give a short explanation of the physical reason for the difference
between the space-time represented by the $y, z$-metrics and the space-time 
given by the $r$-metric. Although all three metrics appear to have 
infinitely large extra dimensions since the extra spatial dimensions run from $y, z, r = - \infty$ to
$y, z, r = + \infty$ only the $y$ and $z$ metric have infinite {\it proper} distance into the bulk. 
For example, consider a path in the $y$ metric \eqref{rsmetric} which goes from $y=0$ to $y=+\infty$
perpendicular to the 4D brane at $y=0$. The proper length of this path, using \eqref{rsmetric}, 
is $s = \int ds = \int _0 ^\infty dy = \infty$. In a similar manner for the $z$ metric one finds that
the proper distance into the bulk for the path $z=0$ to $z=+\infty$ going perpendicular from the
brane is infinite -- $s = \int ds = \int _0 ^\infty \frac{dz}{k |z| +1} = \infty$. However,
for the $r$ metric from \eqref{rsmetric2} one finds that the infinitesimal proper distance for a path 
going from $r=0$ to $r=\infty$ perpendicular to the brane at $r=0$ is $ds = e^{-k |r|} dr$.
Integrating this gives
\begin{equation}
\label{sx}
s= \int ds = \int _0 ^\infty e^{-k r} dr = \frac{1}{k} ~,
\end{equation} 
where in \eqref{sx} we have dropped the absolute value sign since we are integrating over positive $r$.
Note also that the proper distance $s$ in \eqref{sx} would be infinite if one considered 
an increasing warp factor rather than a decreasing warp factor i.e if one let $k \rightarrow -k$. 
For the $y, z$ metrics the proper distance into the bulk is infinite regardless of the choice of the sign
of $k$. Thus the space-time represented by \eqref{rsmetric2} is a single brane at $r=0$ with either
a finite (for $k>0$) or infinite (for $k<0$) proper distance into the bulk dimension. 

In some sense the $r$ coordinate metric of \eqref{rsmetric2} for increasing warp factor
is ``more" of an infinite extra dimension metric that the Randall-Sundrum space-time 
given by \eqref{rsmetric} or \eqref{rsmetric1}. Although the
Randall-Sundrum metrics have an infinite proper distance into the bulk it takes only finite proper time for a
massive particle to ``fall" from the brane at $y, z =0$ to infinity $y, z = \infty$. In \cite{Muck} \cite{Rubakov3}
it was shown that the proper time for a massive particle to fall from the brane to infinity was
$\tau =\pi/2 k$. Following \cite{preston-ajp} one can straightforwardly calculate the proper time 
for a massive particle to ``fall" over the r-metric brane from $r=0$ to $r=\infty$. For an increasing 
warp factor ($k<0$) this yields $\tau = \infty$ while for a decreasing warp factor
($k>0$) this yields $\tau = 1/2 k$, a finite proper time. Thus both in terms of proper distance from $r=0$ to 
$r=\infty$ and in terms of proper time to fall from $r=0$ to $r=\infty$ the $r$ metric of \eqref{rsmetric},
for $k<0$, yields infinite results. One the other hand for a decreasing warp factor, $k>0$, both the proper distance
and proper time are finite.  

\section{Localization of fields of various spins}

We now discuss the localization of fields of different spins (spin $0$, spin $1/2$ and spin $1$) for the 
space-time represented by the metric in \eqref{rsmetric2} i.e. the $r$ coordinate system. As mentioned in 
the Introduction, it is important that most matter fields and fundamental interaction fields, with the possible 
exception of gravity, should be well confined to the 3 + 1-dimensional brane. 
The localization results for spin $0$ \cite{RaSu2, gogber3, gogber4} 
spin $1/2$ \cite{BaGa} and spin $1$ gauge bosons \cite{pomoral} are well known 
for the 5D space-time given by the metrics \eqref{rsmetric} and \eqref{rsmetric1}. The localization of fields of 
various spins is an unresolved issue for the original Randall-Sundrum model. Summarizing briefly the
previous results for localization in the 5D Randall-Sundrum model one finds: (i) spin-$0$ fields are 
localized if $k>0$ (i.e. a decreasing warp factor for our conventions here) but not localized 
if $k<0$ (i.e. an increasing warp factor for our conventions here) \cite{BaGa}; (ii) spin-$1$ fields are not
localized for either $k>0$ or $k<0$ \cite{pomoral}; (iii) spin-$1/2$ fields behave exactly opposite to 
spin-$0$ fields -- they are not localized if $k>0$ (i.e. a decreasing warp factor) 
but localized if $k<0$ (i.e. an increasing warp factor) \cite{BaGa}. One attempt to address
this issue was to consider 6D \cite{gog-sing} \cite{gog-sing1} \cite{Mi} \cite{Mi2} \cite{Mi3}
and higher dimensional brane models \cite{Oda00} \cite{Oda01} \cite{Si}. These higher dimensional 
models did exhibit better localization behavior as compared to the 5D model. 

The localization condition for fields onto the brane is that the action integral over the extra dimension 
should be finite \cite{Si, Mi}. This association between a finite action 
integral and field localization can be considered in terms of the Wick rotated propagator \cite{Zee} 
$\int {Dxe^{ - S\left( x,{it} \right)} }$ or time independent $\int {Dxe^{ - S\left( {x} \right)} }$. 
The propagator and the associated field will vanish unless the action integral is finite. The space-time
background represented by the $r$ coordinates in \eqref{rsmetric2} has
qualitatively different behavior in regard to the localization of fields to the brane as compared to
the space-time background represented by the $y$/$z$ metrics \cite{BaGa, Perez05}.

\subsection{Scalar field}
We first consider a complex scalar field in $r$-metric space-time \eqref{rsmetric2}. 
The action for a complex scalar field can be written as \cite{Davies07, BaGa, Oda01},
\begin{equation}  
\label{PhiAction}
 S_0   =  \int {d^5 x \sqrt g } g^{MN} \partial _M \Phi ^* \,\partial _N \Phi,
\end{equation}
where the subscript $0$ indicates the spin of the field. Notice that we are assuming that the
5D scalar field $\Phi$ is massless in that there is no term like $M^2 \Phi ^* \Phi$.
Applying the principle of least action leads to the equation of motion for the scalar field,
\begin{equation}
\label{scalar-gen}
\partial _M \left( {\sqrt g {\text{ }}g^{MN} \partial _N \Phi } \right) = 0.
\end{equation}
In terms of the general form of the brane metric given in \eqref{gen-metric} the scalar field
equation of motion \eqref{scalar-gen} becomes
\begin{equation}  
\label{PhiEqs2}
\eta ^{\mu \nu } \partial _\mu  \partial _\nu  \Phi  - \frac{1}{{a^2 b}}
\partial _5 \left( {\frac{{a^4 }}{ b }\partial _5 \Phi } \right) = 0,
\end{equation}
where $\sqrt g  = a^4 b$ and $g^{55}  =  - \frac{1}{b^2}$. 
Decomposing the field, $\Phi \left( {x^\mu  ,x^5 } \right) = \varphi \left( {x^\mu  } \right)\chi_0 \left( {x^5 } \right)$, 
the equations for the field on the brane can be written in terms of the separation constant $m$,
\begin{equation}
\label{eta-uv}
\eta ^{\mu \nu } \partial _\mu  \partial _\nu  \varphi  =  - m^2 \varphi.
\end{equation}
The equations of motion for the field in the bulk are as follows,
\begin{equation}
\label{scalar5}
\partial _5^2 \chi _0  + 4\frac{{a'}}{a}\partial _5 \chi_0  - \frac{{b'}}{b}\partial _5 \chi _0  =   
- \frac{b^2}{a^2}m^2 \chi _0.
\end{equation}
Recalling the ansatz functions for the $r$-coordinate metric from \eqref{rsmetric2}
\begin{equation} 
\label{MetricComponents}
a (r)  = b (r)  = e^{ - k\left| r \right| },
\end{equation}
we find the following form of the scalar field equation in the bulk \eqref{scalar5}
\begin{equation}
\label{FieldEqs}
   {\left( {\partial _r^2  - 3k\partial _r  + m^2 } \right)\chi _0  = 0.}  \\
\end{equation}

\subsubsection{Scalar massless modes: $m=0$} 

When $m=0$ the solution to \eqref{FieldEqs} is,
\begin{equation}
\label{scalar-x-m0}
   {\chi _0  (r) = c _0   + b _0 \,e^{3\,kr} ,}
\end{equation}
Previous efforts in studying scalar field zero modes have focused on the constant field
solution which is obtained by setting $b _0=0$. Here we will keep the
non-constant zero mode solution of \eqref{scalar-x-m0} ($b _0 \,e^{3\,kr}$)
since it might have potentially useful localization properties under the change $k \rightarrow -k$ 
i.e. going from a decreasing warp factor to an increasing warp factor.

The action for the scalar field \eqref{PhiAction} in the $r$ metric can be expanded as
\begin{equation} 
\label{BoundInt1}
  S_0   =  \int\limits_0^\infty  {dr \sqrt g } \chi _0 ^* \chi _0 g^{\mu \nu } \int {d^4 x\partial _\mu  \varphi ^* } 
\partial _\nu  \varphi  +  \int\limits_0^\infty  {dr } \sqrt g g^{rr} \partial _r \chi _0 ^{*}  
\partial _r \chi _0 \int {d^4 x\,\varphi ^2 } ~.
\end{equation}
For the scalar field to be localizable means that the two integrals over the bulk dimension $r$  must be finite
i.e. the two integrals
\begin{equation} 
\label{chiInt}
N_0  =  \int\limits_0^\infty  {dr ~ a^2 (r)  b (r) \chi _0 ^* (r)\chi _0(r) ~~~,}~~~~
M_0 ^2  = \int\limits_0^\infty  {dr } ~ \frac{a^2 (r)}{b(r)} \left( \partial _r \chi _0 ^{*}  
\partial _r \chi _0 \right) ,  
\end{equation}
must be finite. Here $N_0$ is the 4-dimensional normalization and $M_0$ is the 4-dimensional mass
of the brane wave function $\varphi$ for the spin-$0$ field. Looking at \eqref{BoundInt1} and taking 
into account that we want $\varphi (x ^\mu)$ to behave like a 4D scalar field we should require that
the finite values of $N_0$ and $M_0$ be 
\begin{equation}
\label{scalar-norm}
N_0=1 ~~~~ {\rm and} ~~~~ M_0=m.
\end{equation}
For the background of the $r$ system of coordinates we have 
$a(r)=b(r)=e^{-kr}$ so the integrals \eqref{chiInt} be written out further as 
\begin{equation} 
\label{chiInt-1}
N_0  =  \int\limits_0^\infty  {dr ~ e^{-3 k r} \chi _0 ^* (r)\chi _0 (r) ~~~,}~~~~
M_0 ^2  = \int\limits_0^\infty  {dr } ~ e^{-kr} \left( \partial _r \chi _0 ^{*} (r) \partial _r \chi _0(r) \right) .  
\end{equation}
In the case of the scalar zero-modes $m=0$ 
and $k>0$ (i.e. a decreasing warp factor according to 
\eqref{rsmetric2}) it is clear that in order for $N_0 , M_0$ to be finite we need to take the constant 
mode in \eqref{scalar-x-m0} by setting $b_0 =0$ so that $\chi (r) = c_0$. With this choice we find 
that the integrals in \eqref{chiInt-1} become $N_0 = \frac{(c_0)^2}{3 k}$ and $M_0 =0$. The result
$M_0=0$ is as expected since we are dealing with the case when the 4D mass
is zero, $m=0$. Also from the first normalization condition in \eqref{scalar-norm} 
we find that we need $c_0 = \sqrt{3 k}$. 

In addition to the localization of the constant scalar field zero modes 
for a decreasing warp factor the $r$-metric also appears to allow one to localize the non-constant zero
modes (i.e. those modes obtained by choosing $c_0=0$ in \eqref{scalar-x-m0}) for an {\it increasing}
warp factor which is obtained by letting $k \rightarrow -k$ in \eqref{rsmetric2}, \eqref{scalar-x-m0}, and
\eqref{chiInt-1}. Making the change $k \rightarrow -k$ and taking into account that now
$\chi _0 (r) = b_0 e^{-3kr}$ the integrals in \eqref{chiInt-1} become $N_0 = \frac{b_0 ^2}{3k}$ and 
$M_0 ^2 = 9 k b_0^2/5$. However having $M_0 \ne 0$ conflicts with $m=0$ unless  we take $b_0 =0$
i.e. we get rid of the non-constant zero mode. Although in the end the result for the scalar field 
zero modes is the same as for the Randall-Sundrum metric - only the constant zero mode is localized
and only for decreasing warp factor - the fact that the integrals in \eqref{chiInt-1} are finite
for an increasing warp factor with the non-constant zero modes already indicates potentially different and
interesting localization behavior for the $r$-coordinate metric versus the usual Randall-Sundrum
metric of \eqref{rsmetric} or \eqref{rsmetric1}. This is exactly what we find for the non-zero
scalar modes which we study next.

\subsubsection{Scalar massive modes: $m \ne 0$}

We now turn to the  case when $m \ne 0$. It is easy to show that \eqref{FieldEqs} has the 
$m \ne 0$ solution
\begin{equation}
\label{scalar-x-m}
\chi _0  = c _0  \,e^{\frac{3}
{2}\,kr} e^{ - \frac{1}
{2}\,r\sqrt {9\,k^2  - 4\,m^2 } }  +b _0 e^{\frac{3}
{2}\,kr} e^{\frac{1}{2}\,r\,\sqrt {9\,k^2  - 4\,m^2 }}.
\end{equation}
First it is clear that in order for the fields to be localized one must have $\frac{3}{2} |k| > m$ since otherwise the
$\exp[ \pm r \sqrt{9 k^2 - 4 m^2} /2]$ term in $\chi _0$ from \eqref{scalar-x-m} will go from
exponentially increasing/decreasing to oscillating in the $r$ direction. As a result it is easy to see
that $N_0$ and/or $M_0$ in \eqref{chiInt-1} will diverge. Thus we have the interesting result that
scalar particles are only confined to the brane of the $r$-coordinate metric up to some mass, $m$, which is
set by, $k$, the degree of warping of the extra dimension. We have taken the absolute
value of $k$ since while $k$ can be positive or negative, depending if one has a decreasing or
increasing warp factor, the mass $m$ should be positive definite. 

Next for $N_0$ the $e^{-3kr}$ term in the integral coming from the metric will
always cancel the $e^{3kr}$ factor coming from $\chi_0 ^* (r) \chi_0 (r)$. Thus in order to have an overall decreasing
exponential in the integrand of $N_0$ (and therefore a finite integral) we must select the first solution
in \eqref{scalar-x-m} by setting $b_0 =0$ i.e. only the solution 
$$
\chi_0 (r) = c _0  \,e^{\frac{3} {2}\,kr} e^{ - \frac{1}{2}\,r\sqrt {9\,k^2  - 4\,m^2 } }
$$
is localizable. Under these conditions the behavior of $N_0$ is
\begin{equation}
\label{n4d-m}
N_0 = c^2 _0 \int\limits_0^\infty  {dr ~ e^{-r \sqrt{9 k^2 - 4m^2}}} = 
\frac{c^2 _0}{\sqrt{9 k^2 - 4m^2}}
\end{equation}
which again shows the requirement that $\frac{3}{2} |k| > m$ in order to localize these massive 
scalar modes. Also from the first equation in \eqref{scalar-norm} we require $N_0=1$
or $c^2 _0 = \sqrt{9 k^2 - 4m^2}$. Note that since only $k^2$ appears in
the integral \eqref{n4d-m} that $N_0$ will be finite even under the change $k \rightarrow -k$
i.e. from a decreasing warp factor to increasing warp factor. Next if we look at $M_0$ in \eqref{chiInt-1} 
we see that it has the behavior
\begin{eqnarray}
\label{m4d-m}
M_0 ^2 &=& \frac{c^2 _0 \left( 3 k - \sqrt{9 k^2 - 4m^2} \right)^2}{4} 
\int\limits_0^\infty  {dr ~ e^{2 k r} e^{- \sqrt{9 k^2 - 4m^2} r}} \nonumber \\
&=& \frac{ \sqrt{9 k^2 - 4m^2} \left( 3 k - \sqrt{9 k^2 - 4m^2} \right)^2}{4( \sqrt{9 k^2 - 4m^2} - 2 k)} ~.
\end{eqnarray}
When $k>0$ (i.e. a decreasing warp factor) one needs the condition $2k < \sqrt{9 k^2 - 4 m^2}$ or
$\frac{\sqrt{5}}{2} k > m$ in order for the scalar fields to be localized. This condition coming from the
finiteness of $M_0$ is similar to the condition coming from the finiteness of $N_0$ which 
requires $\frac{3}{2} |k| > m$. The condition $\frac{\sqrt{5}}{2} k > m$ slightly lowers the mass scale at which
particles are no longer localized to the brane relative to the condition $\frac{3}{2} |k| > m$. 

We did not write the absolute value in the condition $\frac{\sqrt{5}}{2} k > m$ above since changing from
decreasing to increasing warp factor (i.e. letting $k \rightarrow -k$) gives a different localization
condition as we now show. Under the change $k \rightarrow -k$ \eqref{m4d-m} becomes
\begin{eqnarray}
\label{m4d-m-a}
M_0 ^2 &=& \frac{c^2 _0 \left( 3 k + \sqrt{9 k^2 - 4m^2} \right)^2}{4} 
\int\limits_0^\infty  {dr ~ e^{-2 k r} e^{- \sqrt{9 k^2 - 4m^2} r}} \nonumber \\ 
&=& \frac{ \sqrt{9 k^2 - 4m^2} \left( 3 k + \sqrt{9 k^2 - 4m^2} \right)^2}{4( \sqrt{9 k^2 - 4m^2} + 2 k)}~.
\end{eqnarray}
The integrals $N_0$ and $M_0$ are finite for both decreasing warp factor and
increasing warp factor. Thus we found an interesting distinction between the 5D Randall-Sundrum metric
and the $r$-coordinate metric \eqref{rsmetric2} -- massive scalar modes can be localized to the 
brane for both decreasing and increasing warp factors for the $r$-coordinate metric. 

Another interesting feature is that one can fix the mass $m$ {\it self-consistently} in a manner similar to 
the mass generation mechanism of Nambu and Jona-Lasinio \cite{nambu}. 
In order for \eqref{eta-uv} \eqref{BoundInt1} and \eqref{chiInt} to be consistent with one another 
one needs $m^2 = M_0 ^2 (m, k)$ where $M_0$ is a function of $m$ and $k$. Applying 
$m^2 = M_0 ^2 (m, k)$ to \eqref{m4d-m} and \eqref{m4d-m-a} and solving for $m^2$ yields
\begin{equation}
\label{mass-k}
m^2 =\frac{k^2 (11 \pm \sqrt{13})}{8} ~,
\end{equation}
where the quadratic equations from \eqref{m4d-m} and \eqref{m4d-m-a} yield the 
same two, positive mass solutions. 

\subsection{Vector fields}

Next we turn to a spin-1 vector gauge boson field in the $r$-metric space-time \eqref{rsmetric2}. 
The action for the vector field, $A^M (x^N)$, can be written as \cite{BaGa, Oda01, pomoral},
\begin{equation}  
\label{VecAction}
 S_1   =  -\frac{1}{4} \int {d^5 x \sqrt g } g^{MN} g^{RS} F_{MR} F_{NS},
\end{equation}
where the subscript $1$ again indicates the spin of the field and the 5D Faraday
tensor is defined in the usual way as $F_{MN} = \partial _M A_N - \partial _N A_M$. 
Applying the principle of least action leads to the equation of motion for the vector field,
\begin{equation}
\label{vector-gen}
\frac{1}{\sqrt{g}} \partial _M \left( {\sqrt g ~ g^{MN} g^{RS} F_{NS} } \right) = 0.
\end{equation}
For the 5D vector gauge boson field $A_N (x^M)$ we take the following ansatz:  
$A_{x_5}= A_r = const$. and $A_ \mu (x^M ) = a_\mu (x^\mu) c(x_5) = a_\mu (x^\mu) c(r)$. One can see
that \eqref{vector-gen} has the constant solution 
\begin{equation}
\label{vector-const}
c(r) = c_1 ~~ {\rm and} ~~ \partial ^\mu f_{\mu \nu} =0 
\end{equation}
where $f_{\mu \nu} = \partial _\mu a_\nu - \partial _\nu a_\mu$ is the 4D Faraday tensor, and the
last equation in \eqref{vector-const} are the 4D vacuum Maxwell equation. Unlike the 5D Randall-Sundrum
metric which only has a constant solution \cite{BaGa} \cite{pomoral} the $r$ space-time metric
has the following non-constant solution
\begin{equation}
\label{vector-exp}
c(r) = \frac{c_1}{a(r)} = c_1 e^{k |r|} ~~ {\rm and} ~~ \partial ^\mu f_{\mu \nu} =0 ~,
\end{equation} 
as can be verified by direct substitution into \eqref{vector-gen} (for this solution one also needs 
to use the 4D gauge freedom and impose the 4D Lorentz gauge $\partial _\mu a^\mu (x ^\nu) = 0$). The existence of
the non-constant solution \eqref{vector-exp} has interesting consequences for the localization of the gauge boson
to the brane. 

In regard to localization we first look at the constant solution from \eqref{vector-const}. Using the fact
that $\sqrt{g} g^{MN} g^{RS} = a(r) \eta^{MN} \eta^{RS}$ the action in \eqref{VecAction} reduces to
\begin{equation}  
\label{VecAction-const}
 S_1   =  -\frac{c_1 ^2}{4} \int _0 ^\infty dr ~ e^{-kr}
\int d^4 x  ~ \eta^{\mu \nu} \eta^{\rho \sigma} f_{\mu \nu} f_{\rho \sigma} ~.
\end{equation}
In order for the gauge boson to be localized to the brane at $r=0$ one needs the integral over the 
fifth coordinate $r$ to be finite and it is easy to see that for the decreasing warp factor this is
in fact the case $\int _0 ^\infty dr e^{-kr} = 1/k$. The reason for this improved localization of gauge
bosons relative to the Randall-Sundrum metric \eqref{rsmetric} or \eqref{rsmetric1}, is that the $r$-metric
\eqref{rsmetric2} has an ``extra" non-trivial metric component $\sqrt{g_{55}} = e^{-k|r|}$ which shows up in the 
integrand. This extra metric component makes the integral finite for the decreasing warp factor $a(r) = e^{-k|r|}$.
However, for the increasing warp factor $a(r) = e^{k|r|}$ the integral over $dr$ in \eqref{VecAction-const} is
infinite and the spin-1 gauge boson is not localized. This better localization behavior of spin-1 gauge
boson for the constant solution \eqref{vector-const} and decreasing warp factor is not such a big surprise
since as pointed out in \eqref{sx} the proper distance into the bulk is finite for decreasing warp factor.
However the localization of spin 1 fields for decreasing warp factor is not trivial. In the next section we find that
spinor fields are not localizable for the case of a decreasing warp factor i.e. when the proper distance in
the $r$ direction is finite. Thus the spinor case below will show that even if the warp factor is decreasing 
and the proper distance into the bulk is finite it is possible that the field modes can still be non-localizable.
In light of this result for the spinor field the fact that the spin 1 field is localizable for a decreasing warp factor is
a non-trivial result.   

Next we look at the non-constant solution of \eqref{vector-exp}. As before 
$\sqrt{g} g^{MN} g^{RS} = a(r) \eta ^{MN} \eta ^{RS}$. Now however, 
\begin{equation}
\label{emn}
\eta ^{MN} \eta ^{RS} F_{MR} F_{NS} = c^2(r) \eta ^{\mu \nu} \eta ^{\rho \sigma} f_{\mu \rho} f_{\nu \sigma}
+\eta ^{55} \eta ^{\rho \sigma} F_{5 \rho} F_{5 \sigma}.
\end{equation}
This looks promising since the first term contains the terms $c ^2 (r) = e^{2 k |r|}$ which, if one lets
$k \rightarrow -k$ (i.e. go from decreasing warp factor to increasing), can more than compensate for geometric
$a(r)$ term. Specifically under the change $k \rightarrow -k$ the first term on the right hand side
of \eqref{emn} will contribute to the action, $S_1$, from \eqref{VecAction} as
\begin{eqnarray}
\int _0 ^\infty &dr& ~ a(r) c^2 (r) \int d^4 x ~ \eta ^{\mu \nu} \eta ^{\rho \sigma} f_{\mu \rho} f_{\nu \sigma} =
\nonumber \\
 \int _0 ^\infty &dr& ~ e^{-k r} \int d^4 x ~ \eta ^{\mu \nu} \eta ^{\rho \sigma} f_{\mu \rho} f_{\nu \sigma} = 
 \frac{1}{k} \int d^4 x ~ \eta ^{\mu \nu} \eta ^{\rho \sigma} f_{\mu \rho} f_{\nu \sigma}. \nonumber 
\end{eqnarray}
Thus the integral over $dr$ is finite and equal to $1/k$ so this first part of the action reduces to
effective 4D electromagnetism. However as in the scalar case there is an inconsistency due to the
second term in \eqref{emn}. In all of the above we have assumed that 4D Maxwell 
equations are valid i.e. $\partial _\mu f^{\mu \nu} =0$. This equation also implies that the spin-1 gauge 
boson is massless. The second term in \eqref{emn} generates a mass term unless one takes $c_1=0$ in
\eqref{vector-exp}. First we note that $F_{5 \rho} = \partial _5 A_\rho = a_\mu (x^\nu) \partial _r (c(r))$
(recall that we have taken $A_5=A_r =0$). Thus the second term on the right hand side of \eqref{emn}
will give a term equal to
\begin{eqnarray}
\label{emn2}
-  \int _0 ^\infty &dr& ~ a(r) (\partial _r c(r))^2 \int d^4 x a^ \mu (x ^\nu) a _\mu (x^\nu) = 
- \int _0 ^\infty dr~ k^2 e^{-k r} \int d^4 x a^ \mu (x ^\nu) a _\mu (x^\nu) \nonumber \\
= &-& k \int d^4 x a^ \mu (x ^\nu) a _\mu (x^\nu)
\end{eqnarray}
which looks like an imaginary mass term $m = i \sqrt{2k}$ for the spin-1 gauge boson $a^\mu (x^\nu )$. Having 
a mass term (imaginary or real) is inconsistent with our choice $\partial _\mu f^{\mu \nu} =0$. Thus the
choice $\partial _\mu f^{\mu \nu} =0$ forces us to only consider the solution which is constant
in the bulk dimension (i.e. the solution in \eqref{vector-const}) and discard the solution which
is non-constant in the bulk dimension (i.e. the solution in \eqref{vector-exp}). One could try two things
in order to avoid this inconsistency: (i) One could replace the massless Maxwell equations
with the massive Proca equations $\partial _\mu f^{\mu \nu} = -m^2 a_\mu a^\mu$. This might provide
a self-consistent Nambu Jona-Lasinio-like mechanism alternative to the Higgs mechanism for generating 
mass for gauge bosons. (ii) One could start with an initial massive 5D gauge boson instead of a 
massless 5D gauge boson by adding a term $\frac{1}{2} M^2 A_M A^M$ to \eqref{VecAction}. 
We leave these considerations for future investigations. 

\subsection{Spinor Fields}

Finally we consider a 5D spin-1/2 spinor field, $\Psi (x^\mu, r)$, in the $r$-metric space-time \eqref{rsmetric2}. 
The action for  can be written as \cite{BaGa, Oda01},
\begin{equation}  
\label{SpinAction}
 S_{1/2}   = \int {d^5 x \sqrt g } {\bar \Psi} i \Gamma ^M D_M \Psi ~.
\end{equation}
This action leads to the following equation of motion 
\begin{equation}
\label{dirac-5d}
i \Gamma ^M D_M \Psi = (i \Gamma ^\mu D_\mu  + i \Gamma ^r D_r ) \Psi =0 ~, 
\end{equation}
where $\Gamma ^M$  are the curved space-time Dirac gamma matrices which are
related to the Minkowski space-time Dirac matrices, $\gamma ^{\bar M}$, via 
$\Gamma ^M= e^M _{\bar M} \gamma ^{\bar M}$, with $e^{\bar M} _M$ and
$e^M _{\bar M}$ being the funfbein and its inverse respectively. The funfbein 
connects the curved and flat space-times according to $g_{MN} = e^{\bar M} _M
e^{\bar N} _N \eta _{{\bar M} {\bar N}}$ - here barred (unbarred) represented
flat (curved) space-time indices. The covariant derivative is given by
$D_M = \partial_M +\frac{1}{4} \omega ^{{\bar M} {\bar N}} _M \sigma_{{\bar M} {\bar N}}$
where the Dirac tensor is defined via the commutator of Dirac gamma matrices as
$\sigma_{{\bar M} {\bar N}} = \frac{1}{2}[\gamma_{\bar M}, \gamma_{\bar N}]$ 
and the spin connection $\omega ^{{\bar M} {\bar N}} _M$
is given in terms of the funfbein and its derivatives by the following expression     
\begin{eqnarray}
\label{spin-con}
\omega ^{{\bar M} {\bar N}} _M &=& \frac{1}{2} e^{N {\bar M}} (\partial _M e ^{\bar N} _N - \partial _N e ^{\bar N} _M )
-\frac{1}{2} e^{N {\bar N}} (\partial _M e ^{\bar M} _N - \partial _N e ^{\bar M} _M) \nonumber \\
&-& \frac{1}{2} e^{P {\bar M}} e^{Q {\bar N}}( \partial _P e _{Q \bar R} - \partial _Q e _{P \bar R} ) e ^{\bar R} _M ~.
\end{eqnarray}
Calculating the spin connections for the general metric \eqref{gen-metric} obtains the non-zero
elements as
\begin{equation}
\label{spin-con-ab}
\omega ^{{\bar r} {\bar \nu}} _\mu = \delta ^{\bar \nu} _\mu \frac{a'}{b}
\end{equation}
where the primes denote differentiation with respect to $r$.
Specializing to the metric \eqref{rsmetric2} with $a(r)=b(r) = e^{-k |r|}$
one can obtain the covariant derivatives which take the form
\begin{equation}
\label{cov-der}
D_\mu \Psi = \left( \partial _\mu + \frac{1}{2} \frac{a'}{b} \gamma _{\bar r} \gamma_{\bar \mu} \right ) =
\left( \partial _\mu + \frac{i k}{2} \gamma _{\bar \mu} \gamma_{\bar 5} \right ) ~~~;~~~
D_r = \partial _r ~.
\end{equation}
We have replaced $\gamma _{\bar r}$ with the standard 4D $\gamma_{\bar 5}$ taking into account 
the relationship between the two (i.e. $\gamma _{\bar r} = i \gamma _{\bar 5}$ \cite{Rubakov2}) and
we have used $\gamma _{\bar r} \gamma _{\bar \mu} = - \gamma _{\bar \mu} \gamma _{\bar r}$. We now separate 
the 5D spinor as $\Psi (x^M) = \psi (x^\mu) p(r)$ and we also separate $\psi (x^ \mu)$ into left handed and 
right handed spinors as
\begin{equation}
\label{left-right}
\Psi (x ^\mu, r ) = \left( {\begin{array}{*{20}{c}}
   {\psi_R (x_\mu) p_R (r)}  \\
   {\psi_L (x_\mu) p_L (r)}  \\
\end{array}} \right)  ~,
\end{equation}
where $\psi _{L}$ and $\psi _{R}$ are the usual two-component left-handed and right-handed
Weyl spinors which  satisfy the chiral conditions 
$\gamma _{\bar 5} \psi _{L} = - \psi _{L}$ and $\gamma _{\bar 5} \psi _{R} = \psi _{R}$. 
Finally we take $\psi _{R,L}$ to satisfy the following massive Dirac equation
$i \gamma ^\mu \partial _\mu \psi _L = m \psi _R$ and $i \gamma ^\mu \partial _\mu \psi _R = m \psi _L$.
Substituting all of this along with the covariant derivatives of \eqref{cov-der} into the 5D Dirac 
equation \eqref{dirac-5d} gives the following equations for $p_R (r)$ and $p_L (r)$
\begin{equation}
\label{dirac-0}
\partial _x p_R (r)  - 2 k p_R (r) =  - m p_R (r)  ~~~ ; ~~~ 
\partial _x p_L (r)  - 2 k p_L (r) = + m p_L (r) ~.
\end{equation}
These equations have the following solution
\begin{equation}
\label{dirac-3}
p_R (r) = c_{1/2} e^{2 k r - m r}  ~~~;~~~ p_L (r) = d_{1/2} e^{2 k r +  m r} ~,
\end{equation}
where $c_{1/2}$ and $d_{1/2}$ are integration constants. Inserting $p_R , p_L$ from \eqref{dirac-3} in the full 
5D spinor from \eqref{left-right} and in turn inserting this into the action \eqref{SpinAction} we find
that the first term (i.e. ${\bar \Psi} i \gamma^{\bar \mu} \partial_\mu \Psi$ the 4D kinetic energy term) is
\begin{equation}
\label{ke-term}
c^2 _{1/2} \int _0 ^\infty e^{- 2 m r} dr \int d^4 x {\bar \psi}_R i \gamma ^\mu 
\partial _\mu \psi _R + 
d^2 _{1/2} \int _0 ^\infty e^{2 m r} dr \int d^4 x {\bar \psi}_L i \gamma ^\mu 
\partial _\mu \psi _L ~.
\end{equation}
In \eqref{ke-term} the geometric factor $e^{-4 kr}$ is exactly canceled by the $e^{4kr}$ coming 
from $p_{R, L} ^2$. It is easy to see that while the first integral over $r$ is convergent and so it
would appear that one could localize the right handed spinors, $\psi _R$, 
the second integral is divergent and thus at least the left handed spinor, $\psi _L$ are
not localizable. Next turning the remaining two terms from the spinor action
\eqref{SpinAction} which contain $\gamma _{\bar r} = i \gamma _{\bar 5}$ 
(i.e. ${\bar \Psi} \gamma _{\bar 5} (- 2 k  + \partial _r ) \Psi$) we find 
\begin{equation}
\label{int-term}
\int d^5 x {\bar \Psi} \gamma _{\bar 5} (- 2 k  + \partial _x ) \Psi = 
c_{1/2} d_{1/2} \int _0 ^\infty dr \int d^4 x (\psi _L ^{\dagger} \psi _R + \psi _R ^{\dagger} \psi _L ) ~,
\end{equation}
which diverges for both left and right handed fields since $\int _0 ^\infty dr \rightarrow \infty$.
Thus the integral over the extra dimension $r$ for the 5D spinor action \eqref{SpinAction} diverges
for both right handed and left handed fields. Note that in \eqref{int-term} one connects or mixes the
right and left handed spinors which results in the $e^{-mr}$ factor of $p_R$ canceling the
$e^{mr}$ factor of $p_L$. The reason for this mixing is that in the chiral representation we use
while $\gamma _5$ is diagonal, $\gamma _0$, is ``anti"-diagonal so that $\Psi ^{\dag} \gamma_0$
moves the left handed components, $\psi _L p_L$, to be the first two-components of $\Psi$ and
moves the right handed components, $\psi _R p_R$, to be the first two-components of $\Psi$.
The above results agree with the calculations in \cite{Oda01} in the massless limit, but appear to
disagree in the massive limit. In the work \cite{Oda01} it is found that if $m \ne 0$ then the spinor fields
are localized. The apparent difference in results comes first because here we are adding a
4D mass term while in \cite{Oda01} a 5D ``mass" term is added. Further in \cite{Oda01} the 5D mass term
added is not of the canonical form $-m {\bar \Psi} \Psi$ but rather $ i m {\bar \Psi} \Psi$. It is 
because the mass term added is imaginary that results in the localization of the spinor fields.
Thus for our $r$-metric \eqref{rsmetric2} spinor fields are not localized for either increasing or
decreasing warp factors. One could localize the spinor fields by introducing a Yukawa coupling to the
scalar field of the form 
\begin{equation}
\label{yukawa}
g {\bar \Psi} \Psi \Phi ~.
\end{equation}
This type of localization method was first suggested in \cite{Rubakov2} and has been extensively used 
to localize otherwise non-localized fermions to the brane. 

Although introducing a non-gravitational, Yukawa interaction between the 5D spinors and the 5D scalar
in order to localize the spinors fields to the brane at $r=0$, spoils the simplicity of the
present brane model one might think to turn this apparent disadvantage into a positive -- one could address
the fermion family puzzle (i.e. why there appear to be three copies or three families of fundamental fermions) 
using the Yukawa coupling of \eqref{yukawa}. There have been various attempts to address the fermion 
family puzzle using brane worlds \cite{liu07, liu07a, liu09, kodama09}. In particular attempts have been
made to obtain a realistic mass spectrum and CKM matrix elements which mix fermions of different families
\cite{gogber07, aguilar06, kaplan12, guo09}.   

\section{Discussion and conclusions}

In this paper we examined in detail the structure of three different 5D brane metrics given
by \eqref{rsmetric}, \eqref{rsmetric1}, and \eqref{rsmetric2}. By investigating these three metrics in terms
of their associated energy momentum tensors and their invariants like the Ricci Scalars given in 
\eqref{ricciscalary}, \eqref{ricciscalarz}, and \eqref{ricciscalarx} we found that the $y$ and $z$ 
coordinate metrics represented the same space-time -- as was already known --
but that the $r$ coordinate metric \eqref{rsmetric2}, although having a form which
appeared to be some combination of the $y$ and $z$ metrics, in fact represented a different space-time.
As far as we found this difference between the space-time represented by the $y$ and $z$ metrics of
\eqref{rsmetric} and \eqref{rsmetric1} and the space-time represented by the $r$ metric of 
\eqref{rsmetric2} has not been discussed before in the literature.  Unlike the space-time 
associated with the $y$ and $z$ metrics, the $r$ metric of \eqref{rsmetric2} could be either an infinite 
extra dimensions (if the warp factor was increasing i.e. $e^{2 k |r|}$ with $k>0$) or a finite extra dimension
(if the warp factor was decreasing i.e. $e^{-2 k |r|}$ with $k>0$). For the decreasing warp factor, although
the coordinate distance $r$ into the bulk extra dimension was infinite, the proper distance was finite
as discussed around equation \eqref{sx}. On the other hand for an increasing warp factor the proper distance 
into the bulk was infinite as in the case of the $y$ and $z$ metrics. Moreover the $r$ - metric with an
increasing warp factor could be considered as a ``more" infinite extra dimensional metric as compared to
the $y, z$ metrics. For the $y, z$ - metrics, even though the proper distance into the bulk was infinite,
it took only a finite proper time $\tau = \pi / 2 k$ for a massive particle to fall from the brane $y, z=0$ to
$y , z =\infty$. However, as shown at the end of section III for the $r$-metric with {\it increasing} warp
factor, the proper distance into the bulk was infinite and the proper time for a massive particle to fall from the brane
at $r=0$ to $r=\infty$ was infinite. 

Next we analyzed the localization properties of fields with spin $0$, $1/2$, and $1$ to the brane at $r=0$
for the $r$-metric. For scalar and spinor fields we studied the localization of both massless and massive modes.
The spin $0$, scalar field case had massless modes localized to the brane for the decreasing warp factor but not for 
increasing warp factor. This was the same as the localization of massless modes in the $y, z$ metrics. 
The massive modes on the other hand could be localized to the $r$-metric brane for either increasing or decreasing
warp factor. This was an improvement over the localization of massive spin $0$ modes for the $y, z$  metric.
In addition the massive mode scalar field had the interesting requirement that for either increasing or decreasing 
warp factor there was some maximum mass beyond which the massive modes would not be localized. This condition
comes from the extra dimensional part of the scalar wave function (i.e. 
$e^{ - \frac{1} {2}\,r\sqrt {9\,k^2  - 4\,m^2 }}$ from \eqref{scalar-x-m}) and implies 
that one needs the scalar field mass to satisfy $3|k|/ 2 > m$ or $\sqrt{5} k / 2 >m$. Another interesting feature of
the massive modes for the scalar fields was that the mass was fixed via a self-consistent condition
reminiscent of the self-consistent mass generation mechanism of \cite{nambu}. Essentially the calculated mass, $M_0$,
from \eqref{m4d-m-a} depended on both $m$ and $k$ and this mass was required to be equal to 
$m$ yielding the self-consistency condition $m^2 = M^2 _0$ giving rise to two masses, $m$ as given in
\eqref{mass-k}. Both masses in \eqref{mass-k} do satisfy the condition that $3|k|/ 2 > m$.
Thus for the increasing warp factor there are two localized modes which are both massive with the
masses given by \eqref{mass-k}. For the decreasing warp factor one has the condition $\sqrt{5} k/2 >m$ 
and thus there are two localized modes: one massless mode and one massive mode given 
by the lower mass in \eqref{mass-k}. 

For the spin 1 gauge bosons the $r$-metric localized the massless modes to the brane for decreasing warp
factor but not for increasing warp factor. This was an improvement over the usual $y, z$ metrics
where spin 1 gauge bosons were not localized either for increasing or decreasing warp factors. As in the
scalar case it might be that considering massive modes would lead to localization for
both types of warp factors. We leave this investigation of the massive spin 1 gauge bosons for future work.
The fact that spin 1 gauge bosons are localized for a decreasing warp factor for
the $r$-metric is in some sense an expected result since for decreasing warp factor the bulk dimension is
not an infinite dimension -- equation \eqref{sx} shows that the proper distance into the bulk for decreasing
warp factor is finite. One other point about the spin 1 gauge boson 
case for the $r$-metric in comparison to the $y, z$ metrics -- the $r$ metric allowed for a
non-constant extra dimensional part for the spin 1 gauge field as given in \eqref{vector-exp}; for the
$y, z$ metrics the only solution for the extra dimensional part of the field was the constant
solution given in \eqref{vector-const} \cite{BaGa}. For the vector field we did not try to introduce a 
4D mass term via $\partial _\mu f^{\mu \nu} = -m^2 a_\mu a^\mu$ nor did we try to add a 5D mass term
via $\frac{1}{2} M^2 A_M A^M$ to \eqref{VecAction}. We leave this possibility for future work , but we note
that these options may allow one to introduce an alternative to the Higgs mechanism for giving mass to 
spin-1 vector gauge bosons.

Finally, in the case of the spinor fields we found that they could not be localized for 
either decreasing warp factor or increasing warp factor. Thus to localize the spinor 
fields one would need to use some non-gravitational interaction as a means of localization such
as the Yukawa coupling between the spinor and scalar fields, given in 
\eqref{yukawa}. The fact that spinors are not localized for either increasing or decreasing 
warp factors does show that the localization of spin 1 fields to the brane for decreasing warp
is not a trivial result.

\vspace{0.1cm}

{\bf Acknowledgments}
PJ was supported in part by Capital One Bank through an endowed professorship. The work of DS
was supported through a 2012-2013 Fulbright Scholars Grant.

\end{document}